\documentclass[preprint,12pt]{elsarticle}

\usepackage{amssymb}
\usepackage{amsmath}
\usepackage{float}
\usepackage[colorlinks=true, allcolors=magenta]{hyperref}

\journal{Nuclear Physics B}

\newcommand{\Cleviosthin}{Clevios-thin}
\newcommand{\Cleviosthick}{Clevios-thick}
\newcommand{\TPBfull}{1,1,4,4-tetraphenyl-1,3-butadiene}

\newcommand{\DS}{DarkSide-20k}

\begin{document}

\begin{frontmatter}



\title{Temperature-Dependent Photoluminescence of PEDOT:PSS for use as Transparent Electrodes in the \DS\ Time Projection Chamber}

\author[a,b]{N.~Swidinsky}
\author[a,b]{E.~Ellingwood}
\author[a,b]{J.~Hucker}
\author[a,b]{P.~Skensved}
\author[a]{P.C.F.~Di~Stefano}
\author[c]{J.~Mason}
\author[c]{M.~Boulay}
\author[a]{A.~Kemp\footnote{Currently located at Department of Physics, University of Oxford, Oxford, OX1~3RH, United Kingdom}}
\author[a,b]{F.~Schuckman}
\author[d]{Y.~Wang}

\affiliation[a]{organization={Department of Phsyics, Engineering Physics and Astronomy, Queen's University},%Department and Organization
            addressline={64 Bader Lane}, 
            city={Kingston},
            postcode={K7L~3N6}, 
            state={ON},
            country={Canada}}
\affiliation[b]{organization={Arthur B. McDonald Canadian Astroparticle Physics Research Institute, Queen's University},%Department and Organization
            addressline={64 Bader Lane}, 
            city={Kingston},
            postcode={K7L~3N6}, 
            state={ON},
            country={Canada}}
\affiliation[c]{organization={Department of Phsyics, Carleton University},%Department and Organization
            addressline={1125 Colonel By Drive}, 
            city={Ottawa},
            postcode={K1S~5B6}, 
            state={ON},
            country={Canada}}
\affiliation[d]{organization={Institute of High Energy Physics, Chinese Academy of Sciences},%Department and Organization
            addressline={19B Yuquan Road, Shijingshan District}, 
            city={Beijing},
            postcode={100049}, 
            %state={},
            country={China}}

%% Abstract
\begin{abstract}

Dual-phase time-projection chambers (TPCs) filled with noble elements are used for particle detection, with many focusing on rare-event searches. These detectors measure two signals: one from scintillation light, and another from drifting ionized electrons. The \DS\ design uses a transparent vessel with external photodetectors. Electrodes, used to drift the electrons, are located between the active medium and the photodetectors, requiring them to be transparent to allow scintillation light to transmit through them. The transparent electrode coating for \DS\ is poly(3,4-ethylenedioxythiophene) poly(styrene sulfonate) (PEDOT:PSS) or Clevios. For rare-event search detectors, the fluorescence of materials that are between the active volume and the photodetectors may lead to backgrounds in the data. Since Clevios is a new material for TPC electrodes, its fluorescent properties need to be characterized to understand their potential impact on backgrounds. Previous studies have indicated that Clevios can fluoresce, with maximal fluorescence produced using UV-excitation. Our study analyzes the fluorescence of Clevios, under UV-excitation, using both time-resolved and spectral techniques between 4~K and 300~K. We find that the fluorescence of Clevios is negligible when compared to the fluorescence of common fluorescent materials used in TPCs, such as \TPBfull\ (TPB).

\end{abstract}


% \begin{highlights}
% \end{highlights}

%% Keywords

\end{frontmatter}

%\linenumbers

%\newcommand{\DS}{DarkSide-20k}
\section{Introduction}

\begin{figure}[H]
    \centering
    \includegraphics[width = 0.6\linewidth]{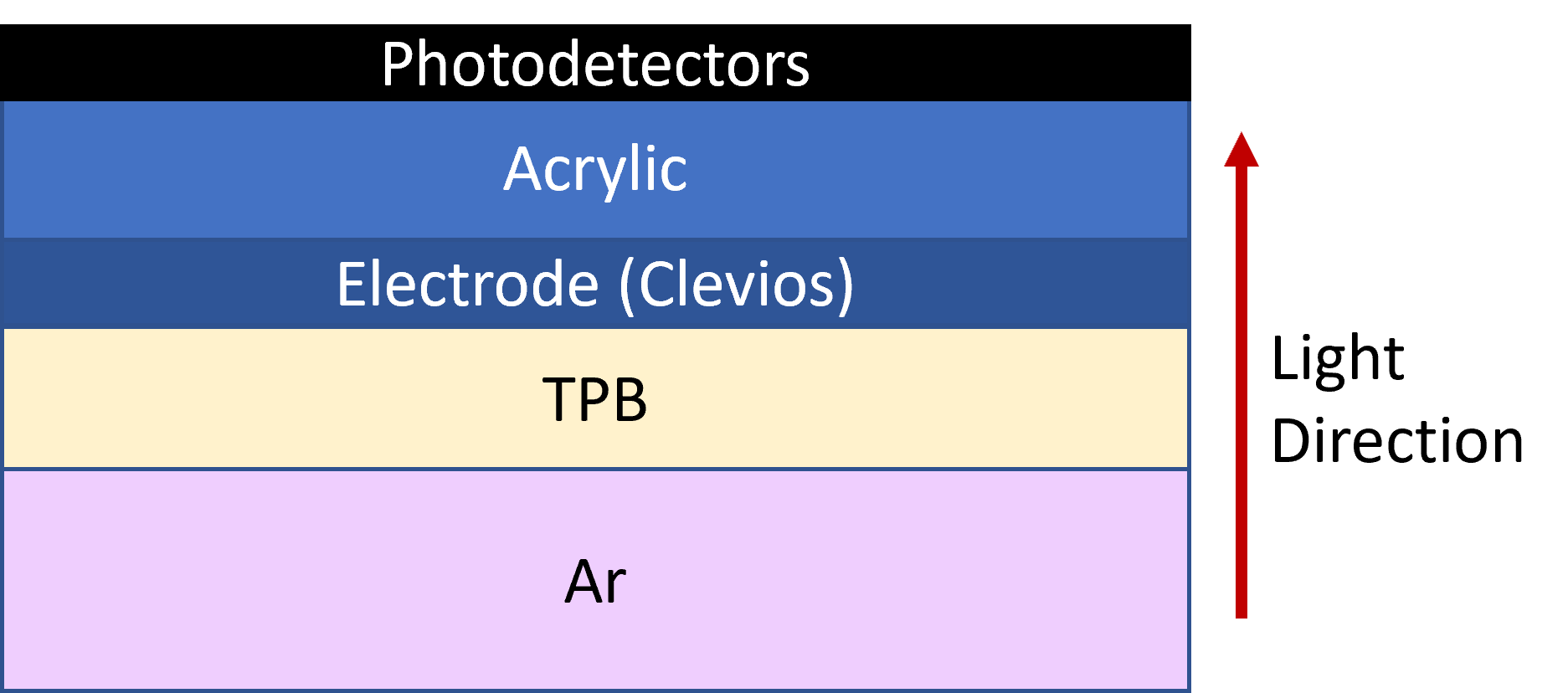}
    \caption{Diagram of the order in which scintillation light from the LAr interacts or passes through various layers before it reaches the  light detectors (SiPMs).  Diagram not to scale.}
    \label{fig:layers}
\end{figure}

Several next-generation noble-liquid dark matter and neutrino detectors, such as \DS~\cite{DarkSide-20K}, DUNE~\cite{DUNE_2021}, XENONnT~\cite{XENONnT_2023}, and LZ~\cite{LZ_overview}, use time-projection chambers (TPCs) to measure particle interactions. In a dual-phase design there are two signals detected per particle interaction. The primary signal comes from scintillation light produced by the particle interaction in the liquid phase which also produces ionization electrons. The second signal is produced when these ionization electrons drift in the TPC electric field and interact with the liquid-gas interface, producing addition scintillation light. Electrodes in the detector create an electric field which drifts the electrons into the gaseous phase, producing a second light signal~\cite{TPC_general}. The \DS\ detector is a dual-phase argon TPC for dark matter detection. Its design uses photodetectors that are outside of the main vessel which requires a transparent vessel and electrodes. TPC electrodes are generally wire grids, but \DS\ uses a conductive coating applied to the inner surface of the vessel instead. In the case of liquid argon (LAr) detectors, such as \DS, the 128~nm scintillation light from events within the LAr first interacts with a wavelength shifter, \TPBfull\ (TPB), producing visible fluorescent light. The wavelength shifter is necessary to shift the ultraviolet (UV) scintillation light to the visible range where most photodetectors, such as silicon photomultipliers (SiPMs), are sensitive. The fluorescent light from the TPB, which peaks around 420~nm, passes through the electrode and the vessel before reaching the layer of photodetectors as shown in Fig.~\ref{fig:layers}. The electrode must be made of a material which is conductive enough to provide the required electric field strength for the TPC while being made of a transparent, and radiopure, material.

Poly(3,4-ethylenedioxythiophene) poly(styrene sulfonate) (PEDOT:PSS), branded as Clevios by Heraeus~\cite{Heraeus_Clevios}, is an organic conductor that is optically transparent when in a thin film. This makes it a candidate for transparent electrodes. Clevios can be applied as a coating to the inside of the transparent vessel, with a TPB coating applied on top of the Clevios. Different thicknesses can be used depending on the visible transparency and resistivity required at the location of the coating.

Previous results~\cite{acrylic_tpb, Corning_2020} have shown that some commercial UV-absorbing acrylics used in other dark matter experiments exhibit a very small amount of fluorescence under UV excitation. As these types of experiments are intended to have high sensitivity and very low backgrounds, fluorescence of the materials used must be mitigated and characterized. Due to its aromatic chemical structure~\cite{pettersson_optical_2002}, Clevios may also fluoresce. Koyama et al.~\cite{koyama_photoluminescence_2015} mapped the emission spectra of PEDOT:PSS under different excitation energies. 
The Clevios coatings in LAr experiments will mostly be excited by the 420~nm (2.95~eV) fluorescence light from the interaction of the 128~nm argon scintillation light with the TPB. The paper shows that 420~nm light will produce little emission in Clevios compared to shorter wavelengths. The maximum fluorescent intensity occurs under UV excitation with 260~nm (4.77~eV), resulting in fluorescence around 350--400~nm (3.6--3.1~eV)~\cite{koyama_photoluminescence_2015}. Clevios is expected to fluoresce very weakly so this study was conducted using excitation around 260~nm to maximize the fluorescence output. We note that the 260~nm wavelength of maximal emission falls in a region of significant absorbance observed in other work on PEDOT:PSS~\cite{bhowal_clevios}.

Our study details the experimental methodology that was employed to compare the amount of fluorescent light produced by Clevios under UV excitation to that of materials of known fluorescence commonly used in the construction of detectors~\cite{acrylic_tpb}. This comparison is done through the detected fluorescence light yield (dLY, the number of photoelectrons for a fixed excitation) of Clevios coated UV-absorbing acrylic as a function of temperature with two different thicknesses of Clevios coatings. These results are compared to the dLY from TPB to produce a relative light yield (rLY) measurement. Additionally, the spectral features of Clevios are studied and compared to the spectral features of the blank substrates without Clevios coatings. This is done to identify unique Clevios fluorescence features in spectra and study how different Clevios coating thicknesses can affect the fluorescence of the sample.

\section{Samples} \label{Sec:samples}

For this study, there were multiple samples used to study the optical properties of Clevios with different combinations of substrates and Clevios coating thicknesses. These are split into samples with an acrylic substrate and those with a fused silica substrate. The UV-absorbing acrylic substrate was used to mimic the conditions of noble-liquid detectors, like \DS, which can be made with an acrylic vessel with coatings on the vessel for the electrode and wavelength shifting. The fused silica substrate was used to measure spectral properties of Clevios independently from the UV-absorbing or fluorescent properties of acrylic. Fused silica is transparent in both the UV and visible region, while the acrylic substrate absorbs light below 375~nm. The high conductivity grade of Clevios used throughout these experiments was Clevios~F~ET. The Clevios coatings are applied by spray coating a solution of Clevios, isopropyl alcohol, ultra-pure water and a surfactant liquid onto the substrates and then  baked in a convection oven at 85${^\circ}$C for 30 minutes to dry. The thickness of the coating on the substrate was determined by the desired sheet resistance for the sample, and the optical transparency. Sample preparation was carried out at Carleton University.

The acrylic substrate is TroGlass Clear Cast 5~mm, a commercial UV-absorbing acrylic from Trotec~\cite{Trotec_acrylic}. 
All of the acrylic substrates come from the same sheet of acrylic, which was machined to fit the dimensions necessary for the cryochamber used for the temperature dependent portion of the study (see Sec.~\ref{sec:spec_cryostat} and Sec.~\ref{sec:time-resolved}). Four samples were prepared on acrylic substrate:  a blank sample with no coating, two samples with Clevios coatings with different thicknesses referred to as \Cleviosthin\ and \Cleviosthick, and one with a 3~$\mu$m TPB coating. The TPB sample is used as a standard reference material to make rLY with Clevios. TPB is an efficient wavelength shifter and will therefore be the largest contributor to the light observed by photodetectors. Fluorescent properties of TPB were studied in more detail in previous measurements~\cite{acrylic_tpb}. Additional samples were prepared using fused silica substrates. The fused silica substrates are Edmund Optics 25x25x2~mm${}^3$, uncoated fused silica windows~\cite{edmund_optics_fs}. Three samples were prepared with this substrate: the blank sample with no additional coating, along with samples with thin and thick Clevios coatings that had approximately the same sheet resistance as \Cleviosthin\ and \Cleviosthick.

The thickness of the Clevios coated samples were measured using stylus profilometry along different parts of the sample as the coating may not be completely uniform across a given sample. Using the stylus profilometry technique on the Clevios coated acrylic substrate samples, the \Cleviosthin\ sample was measured to be $38\pm~3_{stat.}\pm4_{sys.}$~nm while the thick sample was measured to be $173\pm10_{stat.}\pm4_{sys.}$~nm. For the Clevios coated fused silica samples, the thin Clevios coating had a thickness of $49\pm6_{stat.}\pm4_{sys.}$~nm while the thick Clevios coating was $189\pm3_{stat.}\pm4_{sys.}$~nm.

After the samples were coated they were placed in a container which was covered in aluminium foil to protect from UV light, as well as being placed in a nitrogen purged environment to protect against humidity and oxygen, for handling and shipping from Carleton. Upon reception at Queen's University, where this study took place, the samples were kept in that environment until they were placed into the cryostat (a few days to a few weeks after delivery). When the samples were not being used, they were kept in a UV blocking vacuum desiccator. For the absorbance measurements on fused silica, the measurements were taken the day after the samples were received at Queen's to ensure there was limited exposure to humidity, oxygen and UV light.

\section{Absorbance Measurements} \label{sec:absorbance}
\subsection{Absorbance Spectrophotometer}
The absorbance of the samples was measured with an Agilent Cary 60 UV-Vis spectrophotometer. It is capable of measuring the transmittance from 190~nm to 1100~nm using a xenon lamp. Results are presented as absorbance,  defined as $A = -log_{10}T$, where the transmittance of the sample $T = \frac{I}{I_0}$ is the ratio of transmitted to incident light intensities. The samples used in this instrument are the same acrylic and fused silica samples used in other parts of this study with the coating facing the xenon lamp. The device corrects for the contribution of the bare substrate.

\subsection{Absorbance Spectra Results}

\begin{figure}[H]
    \centering
    \includegraphics[width = 0.8\linewidth]{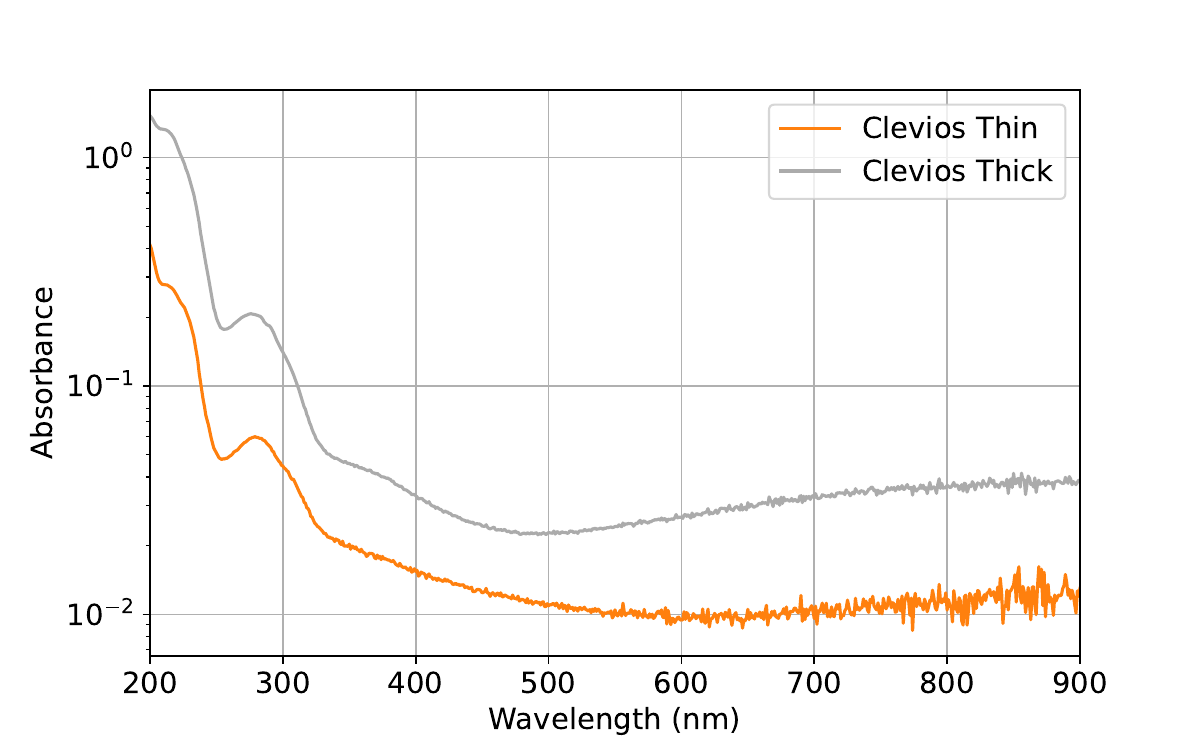}
    \caption{Absorbance of Clevios coatings on a fused silica substrate. The thick Clevios coating has a higher absorbance than the thin coating at all wavelengths, with both coating thicknesses having higher absorbance at lower wavelengths. At lower wavelengths, the local maximum appears at $\sim 275$~nm, and the local minimum at $\sim 255$~nm. }
    \label{fig:FS_absorbance}
\end{figure}

Absorbance measurements were taken for both thicknesses of Clevios on each substrate (see Sec.~\ref{Sec:samples}). Due to the UV absorbing properties of the acrylic substrate, only the absorbance results on the fused silica substrate are used. The absorbance seen in Fig.~\ref{fig:FS_absorbance} shows that the thicker Clevios sample has higher absorbance at all wavelengths as expected.  In addition, both samples have higher absorbance at shorter UV wavelengths than at longer ones. Features seen in the UV region of the absorbance measurements match features seen in previous studies~\cite{koyama_photoluminescence_2015,pettersson_optical_2002}. If experiments apply TPB and Clevios coatings to their vessel as shown in Fig.~\ref{fig:layers}, most of the light reaching the Clevios will be the visible TPB fluorescence around $\sim$420~nm. Based on Fig.~\ref{fig:FS_absorbance}, the absorbance of UV light is higher than visible light, and could lead to more fluorescence. 

The emission spectra that are discussed in Sec.~\ref{sec:emission_spectra} use an excitation wavelength of 260~nm and the time-resolved measurements discussed in Sec.~\ref{sec:time-resolved} use an excitation wavelength of 267~nm. The absorbance values at 260~nm in Fig.~\ref{fig:FS_absorbance} correspond to a transmittance of $(88\pm 3)\%$ for the thin Clevios coating, and $(65 \pm 3)\%$ for the thick one.

\section{Emission Spectra}
\label{sec:emission_spectra}
Emission spectra were measured using two different techniques: measurements with the optical fibre and LED on opposite sides of the sample called transmission measurements, and measurements where the optical fibre and LED are on the same face of the sample called same-side measurements. The transmission measurements were done to probe how the spectrum changes with temperature, while the same-side measurements were done to study spectral features across a broader wavelength range. 

\subsection{Cryostat Measurements} \label{sec:spec_cryostat}
\subsubsection{Cryostat Setup}
\label{sec:cryostat_spectra_setup}

Emission spectra and time-resolved measurements were conducted in an optical cryostat~\cite{acrylic_tpb} that allowed for investigation of fluorescent properties of the samples under a controlled environment. During measurements, the sample is in a vacuum lower than $10^{-6}$~mbar and surrounded by three cryostat shrouds, thermally isolating the sample and mitigating humidity and oxygen concerns. The sample is mounted directly onto the gold-plated copper coldfinger of the cryostat allowing for efficient thermal conductivity between the sample and the coldfinger. The temperature of the cryostat is maintained by a resistive heater operated by an external controller. During measurements, the cryostat temperature is stable within 0.1~K of the desired temperature.

\begin{figure}[H]
    \centering

    \includegraphics[width = 0.8\linewidth]{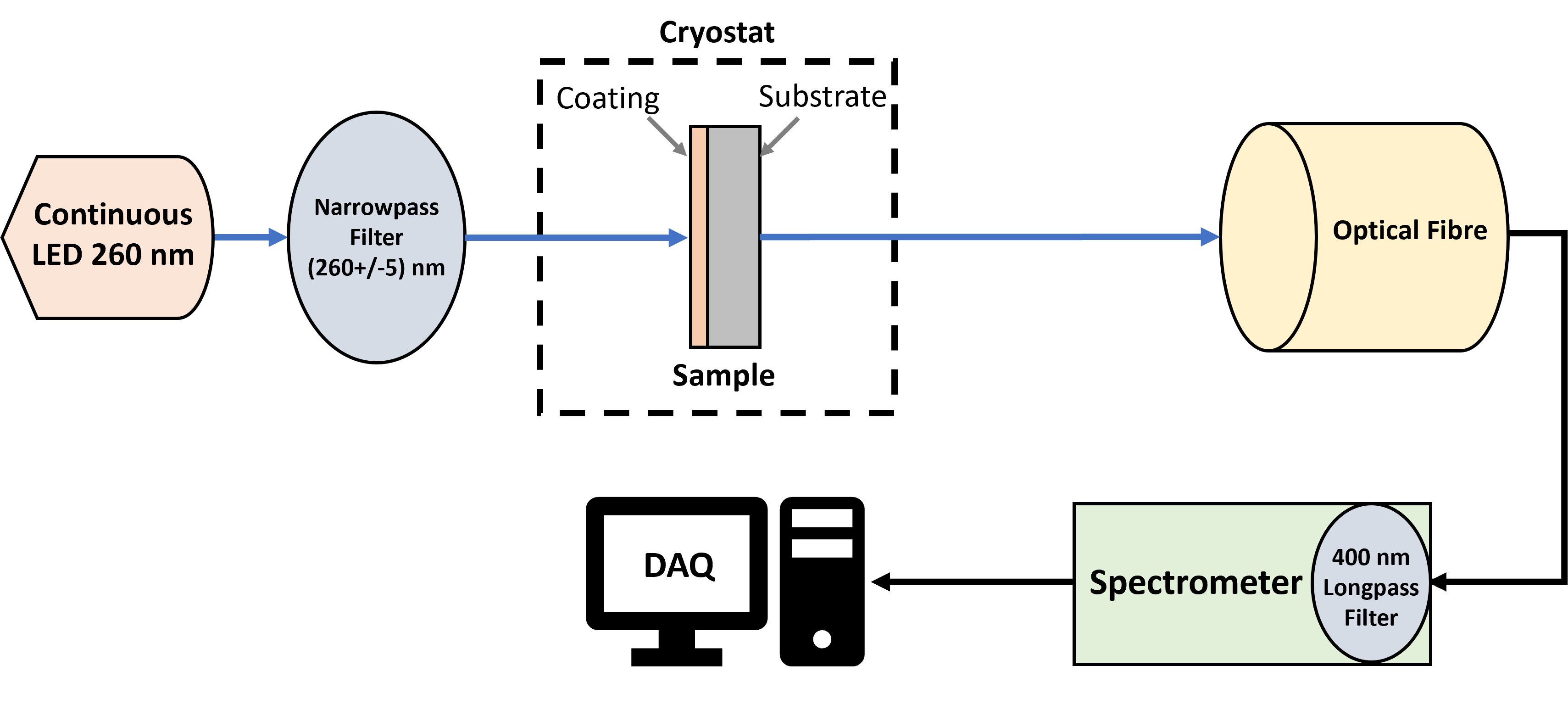}

    \caption{Diagram of the cryostat configuration used for emission spectrometer measurements.}
    \label{fig:cryostat_setup}
\end{figure}

The setup for emission spectra measurements is shown in Fig.~\ref{fig:cryostat_setup}. The geometry of the optical cavity is compact, enhancing light collection but restricting sample geometry as well as beam and photodetector orientation. The sample is excited with a continuous emission 260~nm LED at room temperature attached near one of the outside windows. This wavelength was chosen as it stimulates maximal emission~\cite{koyama_photoluminescence_2015}. This UV  light passes through a 260~nm narrowpass filter before entering the cryostat through borosilicate glass windows in the cryostat shrouds. The light then interacts with the sample coating side first, if a coating is present. Light produced from the sample interaction, such as fluorescent light, passes through the substrate and exits the cryostat via a window opposite the window that the LED was shone through. The light is collected by a UV/Vis/IR optical fibre leading to a Horiba spectrometer, which includes a Symphony II open electrode CCD detector. The optical fibre module has been fitted with a sleeve that is able to be attached directly to the cryostat window. The optical fibre directs the light emitted from the sample to the spectrometer, which is controlled by the data acquisition (DAQ) system. A 400~nm longpass filter in the spectrometer removes higher-order harmonic terms in the spectra from the LED. This filter may also remove emission light at wavelengths between the acrylic absorbance limit (375~nm) and the filter cut-on wavelength (400~nm). Emission spectra were taken with a 10~s exposure at three different temperatures: 300~K, 87~K and 4~K ($\pm 0.1$~K).

\subsubsection{Cryostat Emission Spectra} \label{Sec:Spec_cryostat}

\begin{figure}[H]
    \centering
    \includegraphics[width = 0.9\linewidth]{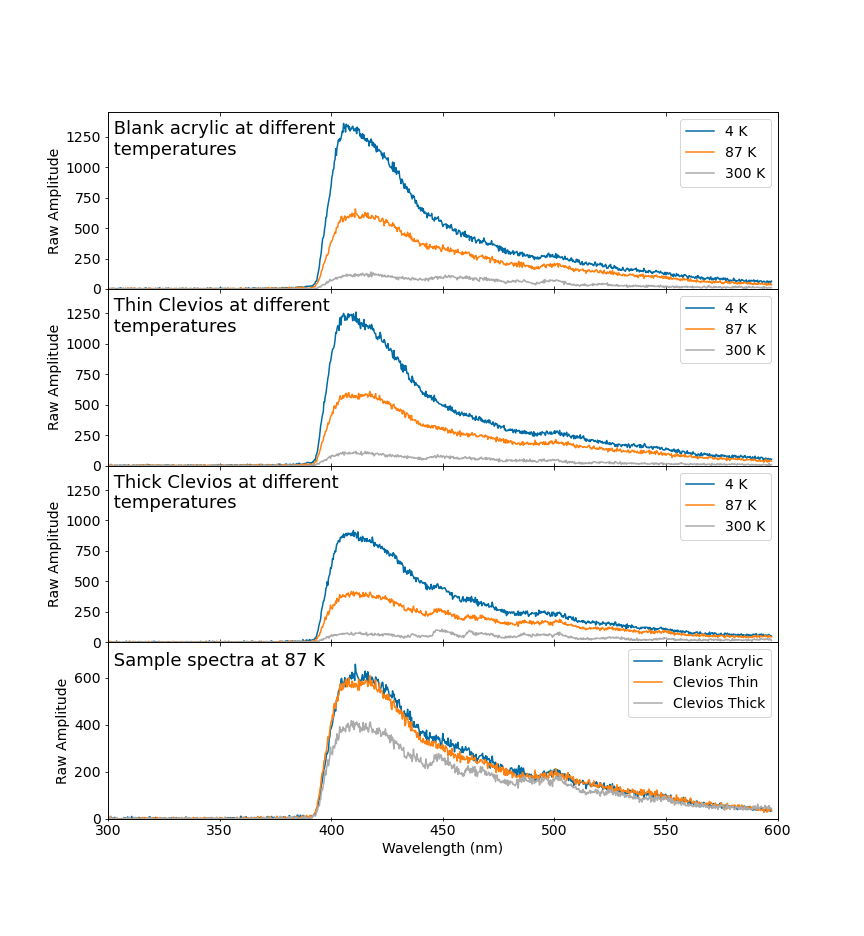}
    \caption{Emission spectra for both Clevios samples compared the acrylic substrate for 10~s exposures. All spectra were taken with a 400~nm longpass filter. Samples show an increase in magnitude as temperature is lowered. Spectral features are similar between the three samples demonstrating that the emission is dominated by the acrylic fluorescence.} 
    \label{fig:Clevios_emission_spec_comp}
\end{figure}

The spectra of the two thicknesses of Clevios were compared to the emission of the acrylic substrate at each of the three temperatures previously mentioned. Figure~\ref{fig:Clevios_emission_spec_comp} shows the spectra of the three samples at the different temperatures in the top three plots as well as a plot comparing the three samples at 87~K at the bottom of the figure. As observed previously~\cite{acrylic_tpb}, the fluorescence of acrylic increases as temperature decreases. Some materials can show different spectral features depending on the temperature, but that was not observed with the blank acrylic substrate or the two Clevios coatings.  The bottom plot of Fig.~\ref{fig:combined_reflection_mode_spectra} depicts the same 87~K spectra from the sample plots above it, but combined to highlight the spectral features and total light output. At 87~K there does not appear to be distinct features that would indicate a strong unique fluorescence from Clevios in the visible range as the spectral features appear in all of the samples including the blank acrylic with no coating. In this plot \Cleviosthin\ has a minimal light output difference in comparison to the blank acrylic sample while \Cleviosthick\ has noticeably less emission. This could be related to the higher absorbance observed for the \Cleviosthick\ coating. It is noted that there may be additional light output between 375~nm and 400~nm that was cut out of the plot because the 400~nm longpass filter was used to block the LED peak at 260~nm and its second harmonic around 520~nm. Light from this missing wavelength range would contribute to the light yields given in the time-resolved experiments.

\subsection{Same-side Measurements}
The previously discussed cryostat emission spectra measure the light that is transmitted through the sample. With UV-absorbing acrylic, any potential UV emission from Clevios would not be observed. 
For this set of measurements, the spectra were taken with the excitation source and the fibre both facing the coating to reduce the effect of the absorbance of the acrylic substrate. 
If Clevios fluoresces isotropically, some of the fluorescence will exit the sample on the same face as the input light and reach the PMT. This configuration thus allows lower wavelength light ($<$375~nm) to be detected. These measurements were carried out at room temperature outside of the cryostat since the cryostat geometry did not allow for desired light source--sample--photodetector configuration.

\subsubsection{Same-side Spectrometer Setup}

\begin{figure}[H]
    \centering
    \includegraphics[width = 0.65\linewidth]{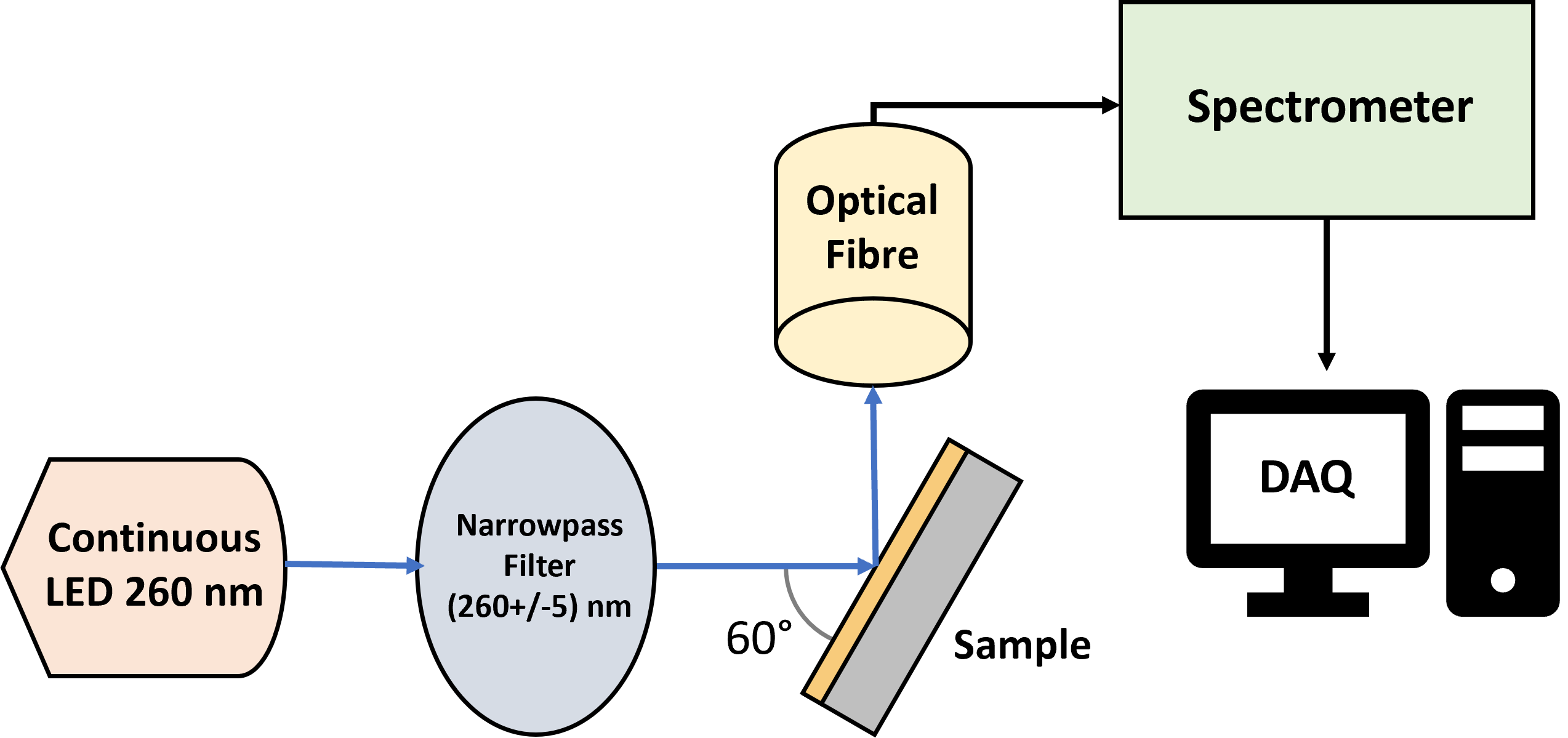}
    \caption{Diagram for the setup in same-side mode. This mode has the LED and optical fibre at 90$^\circ$ to each other and the sample positioned such that the plane of the sample is at a 60$^\circ$ angle to the LED.}
    \label{fig:reflection_mode_setup}
\end{figure}

The setup, shown in Fig.~\ref{fig:reflection_mode_setup}, consists of two arms, which can rotate about a common axis like clock hands. The LED is positioned along one of these arms and the other houses the optical fibre to the spectrometer. The sample is placed on top of the axis and can also be rotated, which allows for different configurations of the setup. The same-side mode is a configuration where the LED and spectrometer are perpendicular to each other and the sample is positioned at an angle compared to the LED to study the emission on the same face of the sample as the coating. The sample is typically placed at an angle of 60${}^{\circ}$ to the LED and the optical fibre placed at 90${}^{\circ}$ to the LED so that the optical fibre is looking at the same face of the sample that the LED is exciting. The sample angle of 60${}^{\circ}$ was chosen to reduce the amount of light from the LED that reflected off the sample surface and reached the optical fibre. The entire setup is in a darkbox to eliminate ambient light from reaching the spectrometer or interacting with the sample.

The same 260~nm LED from the cryostat emission spectra is used for these measurements. For the set of measurements using a given substrate, the LED intensity and the spectrometer exposure time are all identical so that the results can be directly compared.

\subsubsection{Same-side Spectrometer Results} \label{Sec:spec_table_top}

\begin{figure}[H]
    \centering
    \includegraphics[width = 0.95\linewidth]{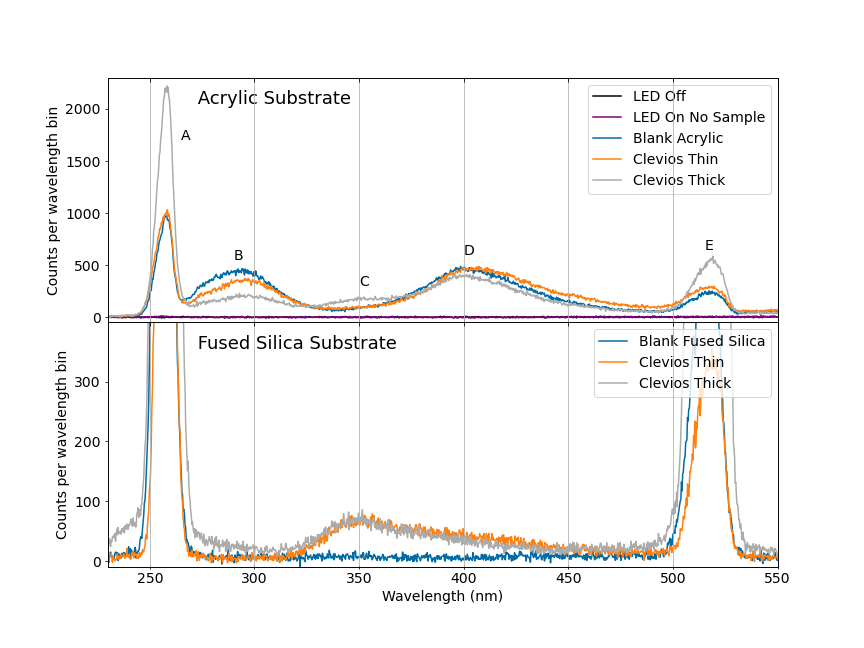}
    \vspace{-0.75cm}
    \caption{Top: Same-face emission spectra of the Clevios coated acrylic samples at room temperature. Features are A: the peak from the LED that is used to excite the sample, B: a potential additive used in acrylic, C: the suspected Clevios fluorescence peak, D: the main acrylic fluorescence peak~\cite{acrylic_tpb} and E: the harmonic of the LED that is a results of the spectrometer respectively. Bottom: Emission spectra of the Clevios coated Fused silica in the same-side setup. The peaks seen in this spectra correspond to the LED at 260~nm, a harmonic of the LED that is produced inside the spectrometer at 520~nm, and a peak at 350~nm, which is associated with the fluorescence of Clevios.}

    \label{fig:combined_reflection_mode_spectra}
\end{figure}

The top half of Fig.~\ref{fig:combined_reflection_mode_spectra} shows the spectra of the samples using acrylic substrate. `LED Off' is a spectrum taken with no LED on to specifically check the ambient light levels in the darkbox. Those results showed that there was no obvious ambient light reaching the spectrometer. The `LED On No Sample' was taken to show that no light is reaching the spectrometer without the sample present. The blank acrylic as well as the two Clevios coatings were tested in this configuration. There are five notable spectral features in Fig.~\ref{fig:combined_reflection_mode_spectra}. Feature A is the peak from the 260~nm LED used for excitation during these measurements, which shows that some of the LED light is reflecting off of the sample and into the spectrometer. Feature A does not have a consistent height between samples, which could indicate that the reflectivity of the Clevios coatings changes with coating thickness. Feature B is suspected to be an additive to the acrylic since this feature is visible in all three samples, but most prominently in the blank acrylic sample with no Clevios coating. Commercial acrylics are not necessarily pure poly(methyl methacrylate) (PMMA) and may have additives for material stability or desired properties such as UV-absorption. Feature C is the fluorescent emission from the Clevios coating. Notably, this feature is only present in the \Cleviosthick\ sample. The absence of this feature in the \Cleviosthin\ spectra could be due to degradation of the Clevios coating. Feature D is the fluorescent emission from the acrylic as seen in Fig.~\ref{fig:Clevios_emission_spec_comp} and previously studied in detail~\cite{acrylic_tpb}. This spectrum shows that for the acrylic-based samples, the overall fluorescence is dominated by the acrylic with, for the thick sample, a small contribution from Clevios peaking around 350~nm.  Finally, feature E is a second-order harmonic of the LED that is created from the diffraction grating within the spectrometer. This feature was eliminated in Fig.~\ref{fig:Clevios_emission_spec_comp} by introducing a 400~nm longpass filter, which prevented the LED light from reaching the diffraction grating and producing harmonics at integer multiples of the LED wavelength.

The bottom half of Fig.~\ref{fig:combined_reflection_mode_spectra} shows the same-side spectra of the samples on a fused silica substrate. This test used a blank fused silica substrate as well as the two thicknesses of Clevios coatings. The resulting spectra has fewer features because the fused silica is not fluorescent when under 260~nm excitation, thus the only features seen in the fused silica blank are the LED reflection peak (aligned with feature A) and the harmonic inside the spectrometer (aligned with feature E). The Clevios samples both show fluorescence at 350~nm, which aligns with feature C in the top half of the plot.

\section{Time-resolved Measurements} \label{sec:time-resolved}
\subsection{Time-resolved Setup}

\begin{figure}[H]
    \centering
        
    \includegraphics[width = 0.8\linewidth]{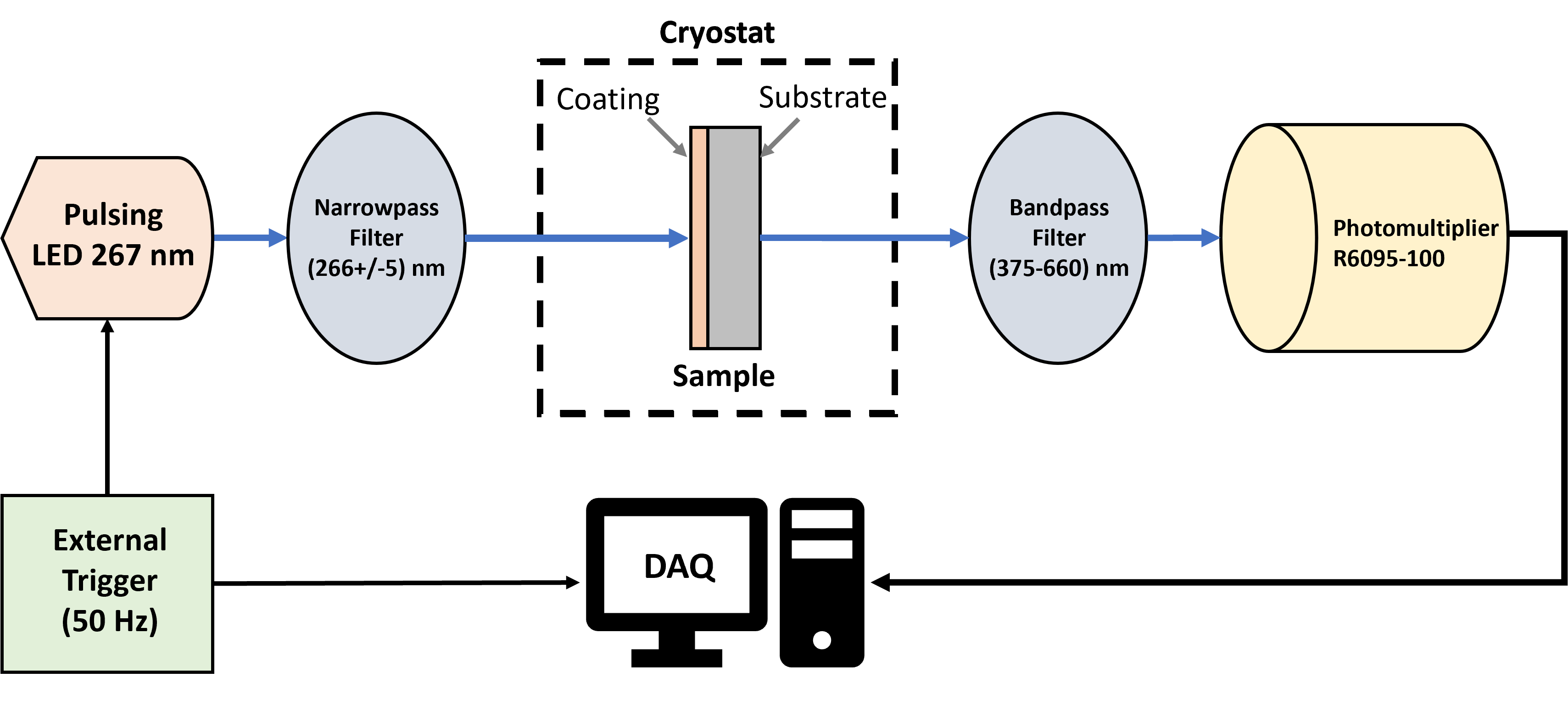}

    \caption{Diagram of the cryostat setup used for time-resolved measurements.}
    \label{fig:cryostat_setup_time}
\end{figure}

Time-resolved measurements are taken in the setup shown in Fig.~\ref{fig:cryostat_setup_time}. The sample orientation within the cryostat is identical to the cryostat emission spectrometer setup discussed in Section~\ref{sec:cryostat_spectra_setup}. For time-resolved measurements, a pulsed 267~nm Horiba Delta Diode LED was used which produced 750~ps long pulses of at a rate of 50~Hz. The LED intensity setting was identical for all samples. Fluorescent light was collected by a Hamamatsu R6095-100~PMT replacing the optical fibre that was used for emission spectra measurements. A function generator was used to produce a 50~Hz pulse which triggered the diode flash and started the PMT readout by the digitizer. The digitizer (10-bit vertical resolution, operating at 2.5~GS/s sampling) wrote the data to disk for further analysis. This data collection process was repeated for 45000 events at every temperature. Each event waveform was 4~$\mu$s long consisting of 10000 samples.

Data is taken at 18 temperatures between 300~K and 4~K, including the boiling points for the noble elements. Between each measurement, the temperature is decreased and then allowed to settle at the desired temperature. The sample was given 30 minutes to thermalize at each temperature before data was taken. This thermalization time was settled on by comparing the LY for different thermalization periods. After the 4~K data was taken the system was heated back to 87~K where cross check measurements were taken to compare the dLY to the earlier 87~K measurement taken when the system was cooled. From 87~K, the cryostat warmed back up to room temperature without the compressor and temperature controller.

\subsection{Analysis}
\label{sec:analysis}

\begin{figure}[H]
    \centering
    \includegraphics[width = 0.8\linewidth]{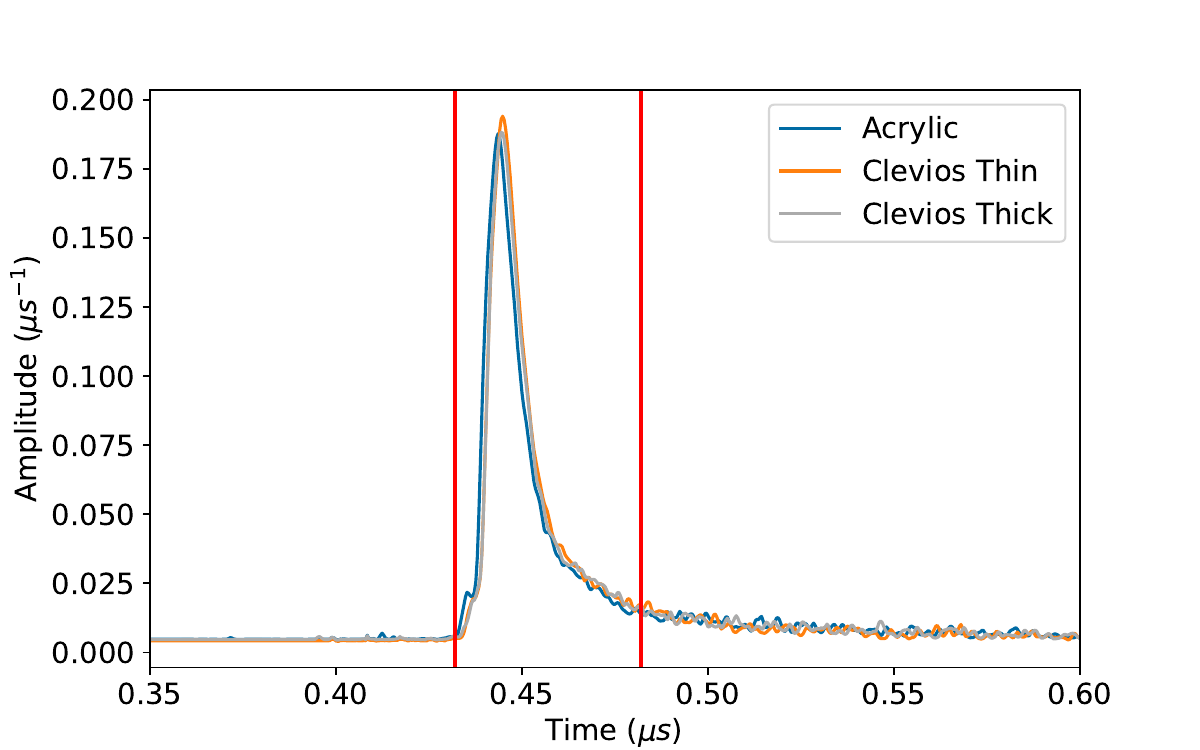}
    \caption{Averaged pulse for bare acrylic, thin Clevios, and thick Clevios.  All pulses have been normalized to an area of 1.  All three pulse shapes are the same, due to the data being instrument response dominated.  Vertical red lines indicate the integration region.}
    \label{fig:average_pulse}
\end{figure}

\begin{figure}[H]
    \centering
    \includegraphics[width = 0.75\linewidth]{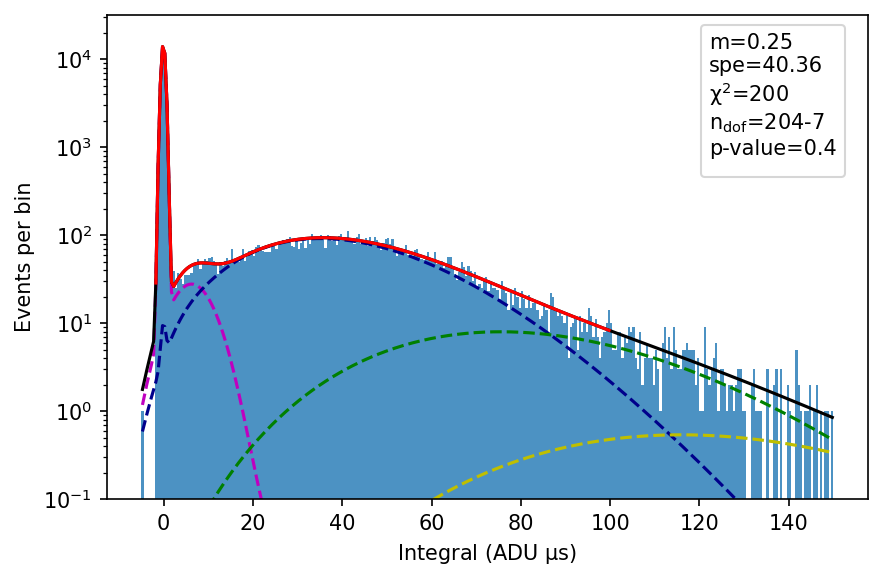}
    \caption{Integral distribution corresponding to the \Cleviosthick~sample fit using the SPE PMT response model.  The pink, dark blue, green, and yellow dashed curves indicate the 0, 1, 2, and 3~PE models, respectively.  The fit was carried out over the range of the red curve, which illustrates the total PMT response model to SPE levels of light.  The black curve is an extrapolation of the red curve using the best fit parameters obtained during the fit and substituting them into the PMT response function. m is is the average number of photoelectrons, spe is the number of ADU $\mu$s/photoelectron.}
    
    \label{fig:int_dist_ex}
\end{figure}

Once the run of 45000 fluorescent events was taken, each event was integrated over a set duration of 50~ns. Averaged pulses for all acrylic substrate samples are shown in Fig.~\ref{fig:average_pulse} along with the integration region. All three pulse shapes are identical to each other, because the measurements are all instrument response dominated. The integration window captures most of the fluorescence pulse, with small amounts of light from the pulse tail incorporated into the systematic uncertainty. The integral represents the charge detected for a specific event in arbitrary units (ADU~$\mu$s). The integrals of all events are placed into a histogram then fitted to determine the average dLY. The excitation light pulse intensity was identical for all samples, but was initially tuned such that the resulting integral distribution for low dLY samples, like acrylic and the Clevios coated acrylic, was a single photoelectron (SPE) distribution. TPB has a much higher fluorescent dLY than acrylic or Clevios so it produces a skewed Gaussian charge distribution. The exact technique for the TPB dLY fitting is described in~\cite{acrylic_tpb}. The integral distribution for Clevios and acrylic, an example of which is provided in Fig.~\ref{fig:int_dist_ex}, was fitted with a model of the statistical processes in a PMT~\cite{tokar_single_1999} returning the average charge integral for one photoelectron, denoted by "spe" in the plot legend, and the average number of SPE per event (ie the dLY), given as "m" in the legend. The fit is shown by the red curve on the plot. 

Statistical errors are determined by the fit of the integral distribution. 
Dominant systematic error comes from the choice of integration window, which was investigated out to $1~\mu s$ to check for additional light. 
Reproducibility of setup in terms of positioning was checked by dismounting and remounting the light detector and light source; this contribution to the systematic uncertainty was found to be negligible.

\subsection{Light Yield Results}

\begin{figure}[H]
    \centering
    \includegraphics[width = 0.8\linewidth]{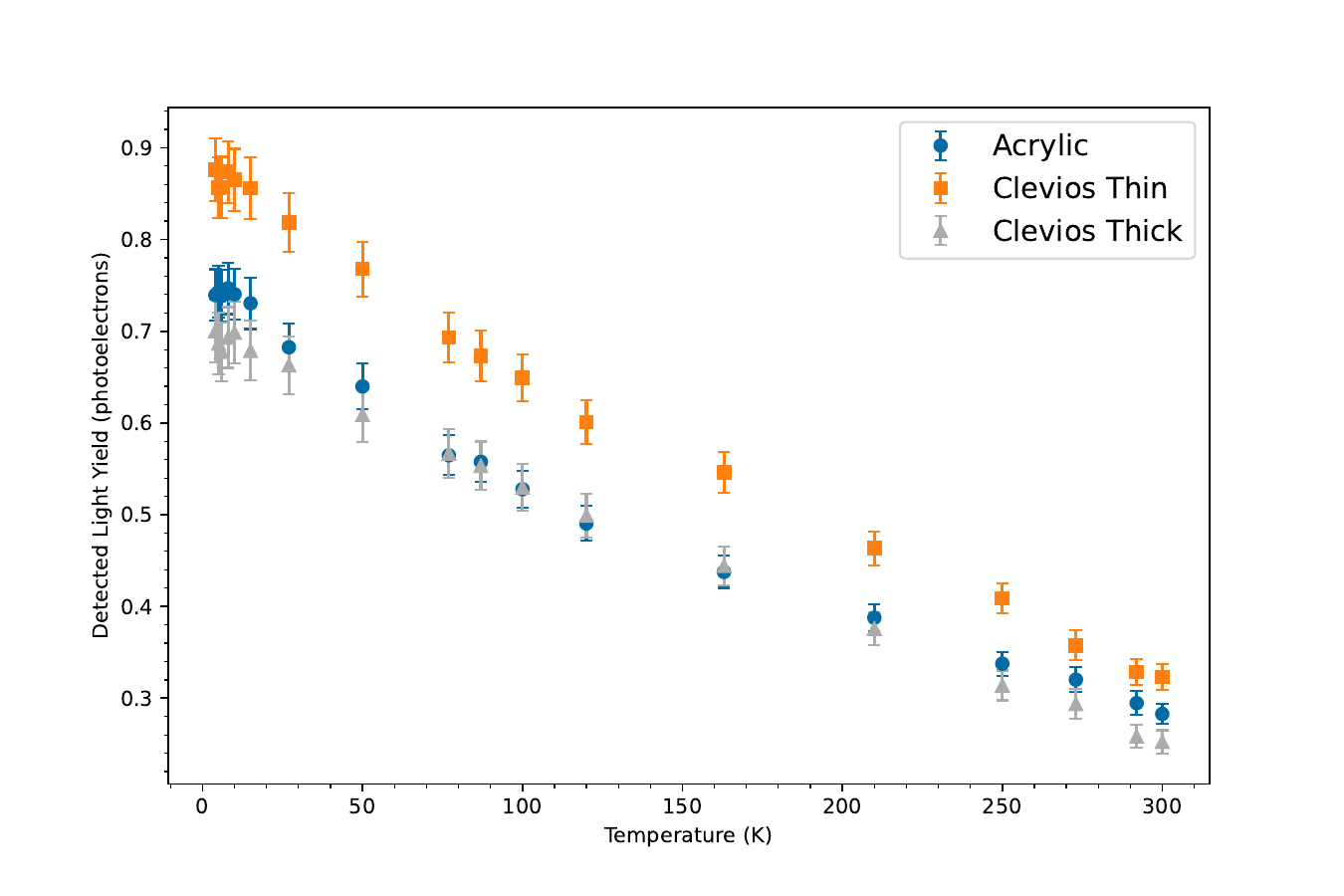}

    \vspace{-1cm}
    
    \includegraphics[width = 0.8\linewidth]{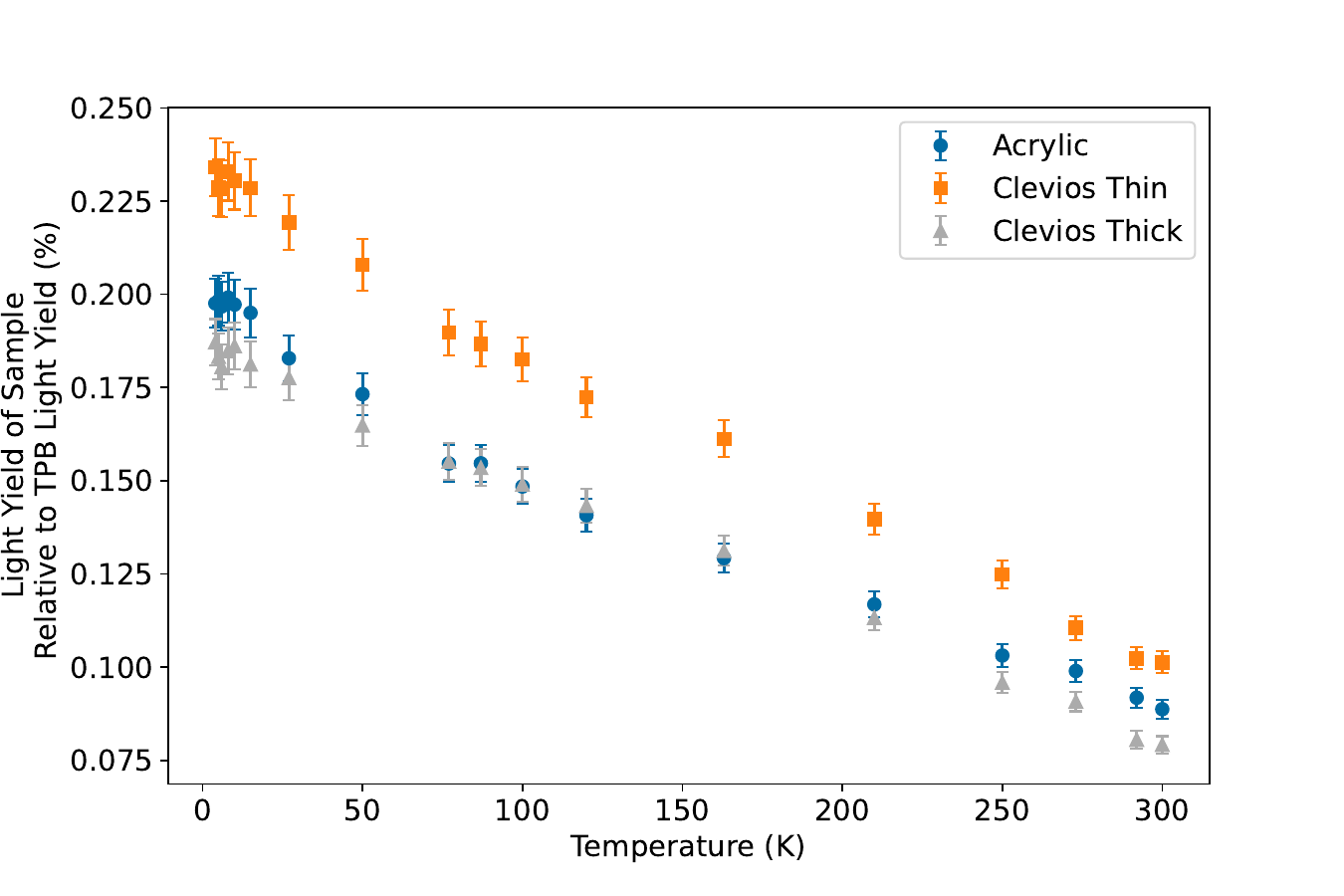}
    \caption{Top : Detected light for both Clevios samples and bare acrylic.  They exhibit a similar dependence on temperature.   Acrylic dominates the light emission.  Bottom: same data, now relative to TPB. Amount of light emitted by Clevios is small compared to that of acrylic, itself small compared to that of TPB. }
    
    \label{fig:acryl_clev100_LYvsT}
\end{figure}

Time-resolved measurements were conducted using both thicknesses of the Clevios-coated acrylic samples. For each temperature, the charge integral of each event is determined following Sec.~\ref{sec:analysis} and the integral distribution is fitted to determine the dLY (Fig.~\ref{fig:int_dist_ex}). Figure~\ref{fig:acryl_clev100_LYvsT} shows that the dLY of the samples increases as temperature decreases, with the \Cleviosthin\ sample producing slightly more fluorescence, and the \Cleviosthick\ sample creating similar fluorescence, to the acrylic sample. The general temperature-dependent trends of the Clevios samples match the trend seen in the acrylic sample confirming that the fluorescence is dominated by the acrylic substrate, itself less than 0.2\% the dLY of TPB at 87~K. Error bars are a combination of statistical and systematic errors discussed in the previous section (Sec.~\ref{sec:analysis}). For use in a TPC, the fluorescence from Clevios is minimal when compared to acrylic and thus should not provide significant background for the detector. 

A detailed explanation of the fluorescence coming directly from Clevios alone, particularly the effect of film thickness, is difficult as there are effects from the combination of the coating and substrate on the light output that cannot be easily decoupled to determine their individual contributions. As shown in Fig.~\ref{fig:combined_reflection_mode_spectra}, the small amount of Clevios fluorescence observed peaks below 375~nm where it will be absorbed by either the acrylic or bandpass filter. Thus, only the tail of the fluorescence at higher wavelengths will contribute to the dLY and rLY observed in Fig.~\ref{fig:acryl_clev100_LYvsT}. The Clevios coating will influence the acrylic fluorescence as incident light may be absorbed or reflected off the Clevios surface, leaving less excitation light to stimulate fluorescence in the acrylic, which makes up the majority of the observed fluorescence. In Sec.~\ref{sec:absorbance}, the thicker Clevios coating had a lower transmittance ($65~\pm~3)\%$ than the thin Clevios ($88~\pm~3)\%$ near the excitation wavelength for these measurements. This could account for the lower dLY and rLY for the thick Clevios coating compared to the thin coating. Thin film interference effects and degradation of the sample due to exposure to the atmosphere and UV light have been considered as additional effects that are influencing the fluorescent light of the samples. Additional data is needed in order to properly characterize these effects and isolate the dLY of Clevios without the substrate.

\clearpage

\section{Conclusion}

We have studied the fluorescent properties of Clevios for use as electrodes in  time projection chambers (TPCs), like \DS, for rare-event searches. In the range of temperatures studied (4~K--300~K), both emission spectroscopy and time-resolved measurements show that the fluorescence of the samples under UV-excitation is dominated by the acrylic substrate with a small effect from the Clevios. These effects include fluorescence from Clevios and other optical properties of Clevios, with transmission in particular having an impact on the fluorescence of the acrylic substrate. The maximum fluorescence of the samples, achieved at low temperature, is less than 0.25\% of the fluorescence of the common wavelength shifter TPB. Emission spectra data show that fluorescence from Clevios peaks at $\sim 350$~nm. Isolating the amount of fluorescence coming from the Clevios requires understanding optical effects such as absorption, reflectivity and interference of the samples, as well as the impact of aging on Clevios. However, for use in TPCs, the fluorescence of Clevios-coated acrylic is minimal, and will have little to no influence on the regular operation of a detector.  

\section{Acknowledgements}

Funding in Canada has been provided by NSERC through SAPPJ grants, by CFI-LOF and ORF-SIF, and by the Arthur B. McDonald Canadian Astroparticle Physics Research Institute.
All of the samples were prepared at Carleton University by Jeff Mason.
Stylus profilometry tests were carried out at NanoFabrication Kingston with the assistance of Dr. Graham Gibson. 
Absorbance measurements were taken using the spectrophotometer at Kevin Stamplecoskie's lab in Queen's University Department of Chemistry
Finally, we thank Guillaume Betrand of CEA Saclay, and Mark Chen, Serge Nagorny, Nir Rotenberg, and James Fraser of Queen's University for providing helpful discussions.

% %% For numbered reference style

%   \textit{\LaTeX: a document preparation system},
%   Addison Wesley, Massachusetts,

% \end{thebibliography}

\bibliographystyle{ieeetr}
\bibliography{references.bib}

\end{document}